# THE GREAT SEASON CLIMATIC OSCILLATION

Ahmed BOUCENNA
Physics Department, Science Faculty
Ferhat Abbas University, 19000 Setif, Algeria
aboucenna@wissal.dz

**Abstract**
The variations of water density and thermal conductivity of the oceans cold region waters according to their salinity lead to suggest an hypothesis of an oscillating climate between two extreme positions: a maximum of hot temperatures and a minimum of cold ones. It will be shown that the distance separating the surface hot streams from the depth cold ones oscillate between two limit values linked to the optimal melting and regeneration glaciers. The melting and regeneration glaciers cycle leads to the Great Saison Climatic Oscillations phenomenon necessary to the regeneration of fresh water resources of our planet.

**1. Introduction**
Is the earth continually getting warmer ? The future of life on our planet depends on the answer given to this very interesting question.
Variations of water density and water thermal conductivity of ocean cold regions according to their salinity leads to thinking about a climate oscillating rather between two extreme positions : maximum hot temperatures and minimum cold ones.
It is known that the earth is subject to various climatic oscillations of relatively short periods such as : the twenty-four hour climatic oscillation period, behind the existence of days and nights, the one-year climatic oscillation period, behind the existence of the four seasons : Spring, Summer, Winter and Autumn/Fall.
The other oscillation that we hypothesize to exist has a longer period, behind the passages of the planet through hot, mild, and cold eras. Our planet lives four Great Seasons: a Great Spring, a Great Summer, a Great Winter and a Great Autumn/Fall, making a Great Year embracing our four small classical yearly seasons.

**2. The Great Season Climatic Oscillation**
The impact of ocean water salinity on the climate has been studied by several scholars[1-7]. Hot streams appear in hot regions of the planet and mild the cold region climate, particularly in the poles. The thermohaline circulation[3,6], for example, is a planet oceanic circulation. This heat circulation through ocean waters is controlled by the sea water salinity[7]. In oceans, the salinities vary from 33 to 37. The average salinity is 34.78. The thermal conductivity K in $Wm^{-1}K^{-1}$ according to the temperature t in °C, the salinity S and the pressure P in dbar, is given, with a good approximation, by:

$$K = 0.5711(1 + 3\ 10^{-3}\ t - 10.25\ 10^{-6}\ t^2 + 6.53\ 10^{-6}\ P - 0.29\ 10^{-3}\ S)$$

Therefore, the higher is salinity of water, the higher is its density and the weaker is its thermal conductivity, and vice versa, that is, the lower is water salinity, lower is its density and the higher is its thermal conductivity.
Supposing that, at given moment, there are in the oceans a hot and low salinity region, and another cold one with a higher salinity due to an optimal quantity of formed glaciers. The hot streams formed of hot and lower density water go on surface, from hot regions to the cold ones to warm them. In cold regions, water becomes colder with high density due to high cold region salinity. Colder water with high density dive deep in oceans giving cold and deep high density streams, which return to hot regions to close the circuit.

A depth d separates surface hot streams going from hot regions to cold ones and deep cold streams moving from cold regions to hot ones.

While surface hot streams arrive in cold region, the glaciers more and more melt and cold waters salinity and density decrease. Therefore, low density cold water dives less deeply in oceans. Lower density cold streams then come progressively and closer to the surface to interfere with surface hot streams. Hot streams effects are then progressively compensated by cold streams. Hot streams are progressively decreased, slowed down or even stopped.

Glaciers regeneration is favoured again. Once more, a deeper cold stream progressively moves away from surface hot streams. Cold streams influence on hot ones decreases. Hot streams arrive again to cold regions provoking again glaciers melting and so forth:

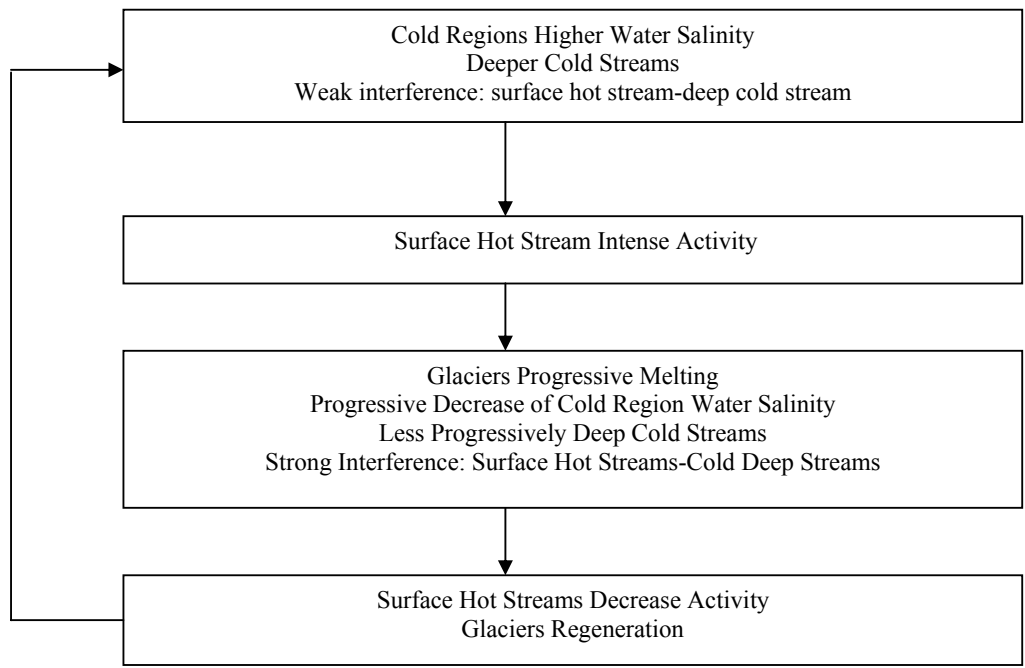

**Great Season Climatic Oscillation cycle**

Depth d separating hot and cold streams is not thus constant. It oscillates between two limit values $d_1$ and $d_2$ according to cold region salinity S that oscillates also between two limit values $S_1$ and $S_2$, corresponding to optimal melting and glaciers regeneration.

This Great Seasons climatic oscillation phenomenon is natural, specific to our planet. It is due to the presence of salt in earth oceans waters.

**3. Great Season Climatic Oscillation Cycle Period**

Thermohaline circulation period is roughly estimated to 1000 years. One could assimilate the Great Season Climatic Oscillation period to this one. To accurately determine the Great Season Climatic Oscillation period, it is essential to know:

       a - Cold streams depth variation according to cold regions salinity:

$$d = f(S)$$

      b - Accurate hot streams movement of the planet.
      c - Melting and regeneration glaciers speed
      d - Other factors influence such as atmosphere $CO_2$ gas concentration and subsequent phenomena.

Having not all necessary data to exactly determine these climatic oscillations period generating Great Seasons, historic based-observations have been used, estimating this period to equal eight centuries (T = 800 to 1000 years). The length of a Great Season would be equivalent to two or two and a half centuries (200 to 250 years).
Culminating Great Summers heat weather maxima would take place around the years $2000 \pm (800$ to $1000)$ k, where k is an integer number:

  -5200, - 4400, -3600, -2800, -2000, -1200, -400, 400, 1200, 2000, 2800, 3600, …

Culminating Great Winters cold weather maxima would be expected to occur around the years $2400 \pm (800$ to $1000)$ k, where k is an integer number:

    - 4000, -3200, -2400, -1600, - 800, 1, 800, 1600, 2400, 3200, 4000, …

The last Great Winter coincides with the mini ice age that Europe experienced, about which contemporary writers have testified. For the time being, a Great Summer is being experienced.

**4. Great Season Climatic Oscillations Consequences**
Great Winter experience makes it possible for earth to regenerate its fresh water resources. Great Winter abundant snows favour regeneration of water underground expanses. So, poor water regions potentials, such as the Maghreb and the Middle East, increase their capacity. Dry rivers, of the last Great Summer, start flowing again. Forest regenerations in terms of population in trees and others are observed. Centre relocation, known to be hot and cold in the planet, provokes particular unaccustomed climatic disturbances. The north habitants spend difficult periods during the Great Winter coldest period, while those living in arid regions live difficult moments during the Great Summer hottest period, with drought and lack of water for people and animals. Some species can even disappear. History, sociology, peoples and migration flux are also influenced by these climatic oscillations.

**5. Conclusion**
Water density and thermal conductivity variation of oceans cold waters regions according to their salinity govern hot stream movements to the poles generating an oscillating climate between two extreme positions: a hot temperature maximum and a cold temperature minimum. The distance separating surface hot streams from deep cold streams oscillates between two limit values linked to optimal melting and glaciers regeneration. The melting and glaciers regeneration oscillations cycle leads to Great Saison Climatic Oscillations phenomenon necessary to fresh water resources regeneration. This phenomenon can be accelerated by the effects of $CO_2$ concentration in atmosphere, due to the pollution.

**References**
1) Eric Guilyardi, La Météorologie - n° 33 - mai 2001, 34-44
2) Broecker W. S., G. Bond et M. Klas, 1990 : A salt oscillator in the glacial atlantic? 1. The concept. *Paleoceanography,* 5, 469-477.
3) Bryan F., 1986 : High-latitude salinity effects and interhemispheric thermohaline circulation. *Nature,* 323, 301-304.
4) Chahine M. T., 1992 : The hydrological cycle and its influence on climate, a review. *Nature,* 359, 373-380.
5) Mikolajewicz U. et E. Maier-Reimer, 1994 : Mixed boundary conditions in OGCM and their influence on the stability of the model's conveyor belt. *J. Geophys. Res.*,
99, 22633-22644.
6) Rahmstorf S., 1996 : On the freshwater forcing and transport of the atlantic thermohaline circulation. *Clim. Dyn.*, 12, 799-811.
7) Vialard J. et P. Delecluse, 1998a : An OGCM study for the TOGA decade. Part I: Role of salinity in the physics of the western Pacific fresh pool. *J. Phys. Oceanogr.*, 28,1071-1088.

**Acknowledgments**
Thanks are due to Professor A. Layadi and Dr S. Keskes for their help.

# L'OSCILLATION CLIMATIQUE DES GRANDES SAISONS

**Ahmed BOUCENNA**
Département de Physique, Faculté des Sciences
Université Ferhat Abbas, 19000 Sétif, Algeria
aboucenna@wissal.dz

**Résume**
Les variations de la densité et de la conductivité thermique des eaux des régions froides des océans en fonction de leur salinité conduisent à suggérer l'hypothèse d'un climat oscillant entre deux positions extrêmes : un maximum de températures chaudes et un minimum de températures froides. On montre que la distance séparant les courants chauds de surface et les courants froids des profondeurs oscille entre deux valeurs limites liées à la fonte et la régénération optimales des glaciers. Le cycle des oscillations de la fonte et de la régénération des glaciers entraîne le phénomène des Oscillations Climatiques des Grandes Saisons nécessaire à la régénération des ressources en eau douce de notre planète.

## 1. Introduction

La terre est-elle entrain de se réchauffer continuellement ? L'avenir de la vie sur notre planète dépend de la réponse donnée à cette question fort intéressante.

Les variations de la densité et de la conductibilité thermique des eaux des régions froides des océans en fonction de leur salinité conduisent à penser plutôt à un climat oscillant entre deux positions extrêmes : un maximum de températures chaudes et un minimum de températures froides.

Nous savons que la terre est sujette à une multitude d'oscillations climatiques de périodes relativement courtes comme : l'oscillation climatique de période vingt quatre heures, qui se traduit par l'existence du jour et de la nuit, l'oscillation climatique de période un an, qui se traduit par l'existence des quatre saisons : hiver, printemps, été et automne.

L'autre oscillation que nous préconisons est de période plus longue. Elle se traduit par le passage de la planète par des ères chaudes, des ères douces, et des ères froides. Notre planète vit quatre Grandes Saisons : un Grand Hiver, un Grand Printemps, un Grand Eté et un Grand Automne constituant une Grande Année qui viennent envelopper nos quatre petites saisons et notre petite année.

## 2. L'Oscillation Climatique des Grandes Saisons

L'impact de la salinité des eaux des océans sur le climat a été étudié par plusieurs auteurs[1-7]. Des courants chauds naissent dans les régions chaudes de la planète et viennent adoucir le climat des régions froides en particulier les pôles.

La circulation thermo haline[3,6], par exemple, est une circulation océanique à l'échelle de la planète. Cette circulation de chaleur à travers les eaux des océans est régulée par la salinité de l'eau des mers[1,7]. Dans les océans les salinités varient de 33 à 37. La salinité moyenne est 34.78.

La conductivité thermique K en $Wm^{-1}K^{-1}$ en fonction de la température t en °C, de la salinité S et de la pression P en dbar est donnée avec une bonne approximation par :

$$K = 0.5711(1 + 3\ 10^{-3}\ t - 10.25\ 10^{-6}\ t^2 + 6.53\ 10^{-6}\ P - 0.29\ 10^{-3}\ S)$$

Ainsi, plus la salinité de l'eau est élevée plus sa densité est élevée plus sa conductibilité thermique est faible, et vice versa, plus la salinité de l'eau est faible plus sa densité est faible, plus sa conductivité thermique est élevée.

Supposons qu'à un instant donné, nous avons dans les océans une région chaude et peu salée, et une autre région froide avec une salinité plus élevée à cause d'une quantité optimale de glaciers formés. Des courants chauds, formés d'eau chaude de moindre densité, vont se diriger, en surface, des régions chaudes vers les régions froides pour les réchauffer. Arrivée en régions froides, l'eau va devenir plus froide et de densité élevée à cause de la salinité plus élevée de la région froide. L'eau plus froide et de densité plus élevée plonge alors profondément dans les océans donnant des courants de densité plus élevée, froids et profonds, qui retournent aux régions chaudes pour boucler le circuit.

Une profondeur d sépare les courants chauds, de surface, se dirigeant des régions chaudes vers les régions froides des courants froids profonds se dirigeant des régions froides vers les régions chaudes.

Au fur et mesure que les courants chauds de surface arrivent en région froide, les glaciers fondent, la salinité et la densité des eaux froides diminuent. L'eau froide de densité moins élevée plonge alors moins profondément dans les océans. Les courants froids, de densité moins élevée, se rapprochent alors progressivement de la surface et interfèrent avec les courants chauds de surface. Les effets des courants chauds sont alors progressivement compensés par les courants froids. Les courants chauds sont progressivement diminués, ralentis, voire arrêtés.

La régénérescence des glaciers est à nouveau favorisée. A nouveau un courant froid de plus en plus profond, s'écarte de plus en plus des courants chauds de surface. L'influence des courant froids sur les courants chauds diminue. Les courants chauds arrivent à nouveau vers les régions froides relançant à nouveau la fonte des glaciers et ainsi de suite :

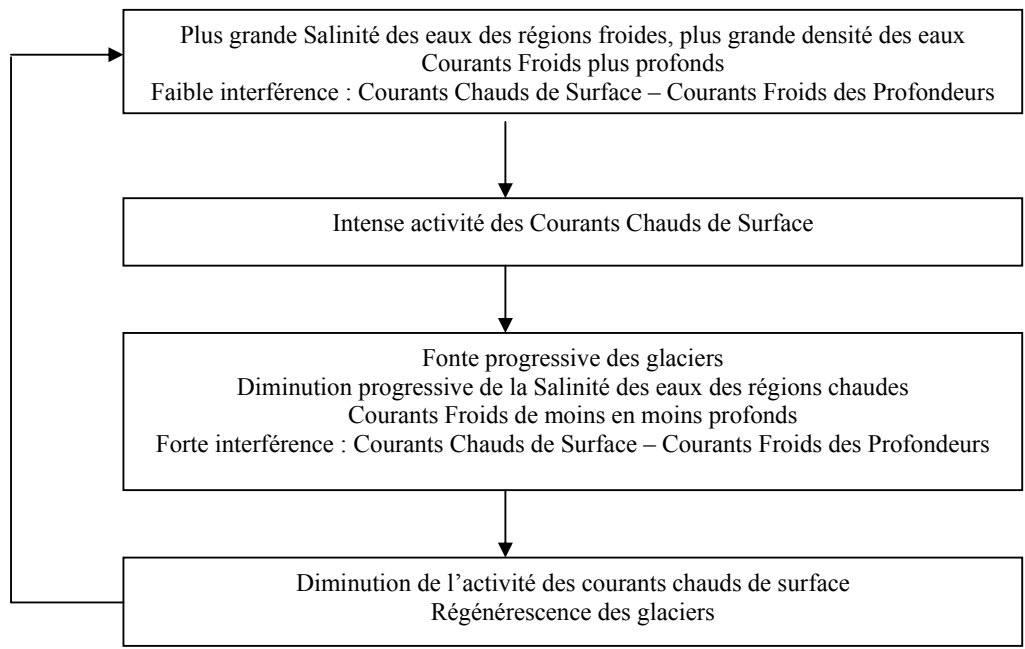

**Cycle des oscillations climatiques des Grandes Saisons**

La profondeur d séparant les courants chauds et froids n'est donc pas constante. Elle oscille entre deux valeurs limites $d_1$ et $d_2$ suivant la salinité S de la région froide qui oscille elle aussi entre deux valeurs $S_1$ et $S_2$ correspondant à la fonte et la régénération optimales des glaciers.

Ce phénomène de l'oscillation climatique des Grandes Saisons est un phénomène naturel, propre à notre planète. Il est dû à la présence des sels dans les eaux des océans.

**3. Période du cycle des oscillations climatiques des Grandes Saisons**

La période de la circulation thermo haline est estimée approximativement à 1000 ans. On pourrait assimiler la période des oscillations climatiques des Grandes Saisons à cette période. Pour déterminer avec précision la période des oscillations climatiques des Grandes Saisons, il est indispensable de connaître :

    a - La variation de la profondeur des courants froids en fonction de la salinité des régions froides :

$$d = f(S)$$

    b - Le mouvement précis des courants chauds et froids de la planète.
    c - La vitesse de fonte et de régénération des glaciers.
    e - L'influence des autres facteurs comme la concentration du gaz $CO_2$ dans l'atmosphère et les phénomènes liés.

N'ayant pas toutes les données nécessaires à la détermination exactes de la période de ces oscillations climatiques engendrant les Grandes Saisons, en se basant uniquement sur des observations historiques la période est estimée approximativement à huit siècles (T = 800 à 1000 ans). La durée d'une grande saison serait de deux à deux siècles et demi (200 à 250 ans).

Les maximums de chaleur, les points culminants des grands étés, seraient situés autour des années 2000 ± (800 à 1000) k, où k est un nombre entier :

   -5200, -4400, -3600, -2800, -2000, -1200, -400, 400, 1200, 2000, 2800, 3600, …

Les maximums de froid, les points culminants des Grands Hivers, seraient situés autour des années 1600 ± (800 à 1000) k, où k est un nombre entier :

    -4000, -3200, -2400, -1600, -800, 1, 800, 1600, 2400, 3200, 4000, …

Le dernier Grand Hiver se confond avec le mini age glacial qu'a traversé l'Europe et dont témoignent des écrivains de l'époque et nous traversons actuellement un Grand Eté.

**4. Conséquences des oscillations climatiques des Grandes Saisons**

Le passage par le Grand Hiver permet à la terre de régénérer ses ressources en eau douces. Les neiges abondantes qui tombent durant le Grand Hiver permettent la régénération des nappes souterraines d'eau. Ainsi les réserves des régions assez pauvres en eau comme le Maghreb et le Moyen Orient, se reconstituent. Les rivières, sèches durant la dernière partie du Grand Eté, se mettent à couler de nouveau. Les forêts régénèrent leur peuplement en arbres et autres.

La délocalisation des centres réputés être chauds et froids de la planète provoque des perturbations climatiques particulières et inhabituelles.

Les habitants du Nord traversent des périodes difficiles durant la partie la plus froide du Grand Hiver et ceux des régions arides traversent leur plus difficiles moments durant la partie la plus chaude du Grand Eté avec la sécheresse et le manque d'eau pour les hommes et pour les animaux. Des espèces peuvent disparaître.

L'histoire, la sociologie et les flux de migration des peuples et des nations sont aussi influencés par ces oscillations climatiques.

## 5. Conclusion

Les variations de la densité et de la conductivité thermique des eaux des régions froides des océans en fonction de leur salinité gouvernent les courants chauds vers les pôles de la terre et engendre, un climat oscillant entre deux positions extrêmes: un maximum de température chaude et un minimum de température froide. La distance séparant les courants chauds de surface et les courants froids des profondeurs oscille entre deux valeurs limites liées à la fonte et la régénération optimales des glaciers. Le cycle des oscillations de la fonte et de la régénération des glaciers entraîne le phénomène des Oscillations Climatiques des Grandes Saisons nécessaire à la régénération des ressources en eau douce de notre planète. Ce phénomène peut être accéléré par les effets de la concentration en $CO_2$ de l'atmosphère liée à la pollution.

**Références**
1) Éric Guilyardi, La Météorologie - n° 33 - mai 2001, 34-44
2) Broecker W. S., G. Bond et M. Klas, 1990 : A salt oscillator in the glacial atlantic? 1. The concept. *Paleoceanography,* 5, 469-477.
3) Bryan F., 1986 : High-latitude salinity effects and interhemispheric thermohaline circulation. *Nature,* 323, 301-304.
4) Chahine M. T., 1992 : The hydrological cycle and its influence on climate, a review. *Nature,* 359, 373-380.
5) Mikolajewicz U. et E. Maier-Reimer, 1994 : Mixed boundary conditions in OGCM and their influence on the stability of the model's conveyor belt. *J. Geophys. Res.*, 99, 22633-22644.
6) Rahmstorf S., 1996 : On the freshwater forcing and transport of the atlantic thermohaline circulation. *Clim. Dyn.*, 12, 799-811.
7) Vialard J. et P. Delecluse, 1998a : An OGCM study for the TOGA decade. Part I: Role of salinity in the physics of the western Pacific fresh pool. *J. Phys. Oceanogr.*, 28,1071-1088.